\documentclass[aps,preprint,nofootinbib,amsmath,11pt]{revtex4}

\usepackage{verbatim}
\usepackage{yfonts}
\usepackage{bm}
\usepackage{epsfig}
\usepackage{psfrag}

\newcommand{\qn}{{\textswab{q}}}
\newcommand{\wn}{{\textswab{w}}}
\def\Go{\gamma_0} 
\def\Gu{\gamma_u} %
\renewcommand{\d}{\partial}

\begin{document}

\title{Quasinormal spectrum and the black hole membrane paradigm}

\author{Andrei Starinets}
\email{starina@perimeterinstitute.ca}
\affiliation{Perimeter Institute for Theoretical Physics,
  Waterloo, ON N2L 2Y5, Canada}

\date{June 2006\\[30pt]}

\begin{abstract}
\vspace{0.5cm}
\noindent
The membrane paradigm approach to black hole physics 
introduces the notion of a stretched horizon as 
a fictitious time-like surface endowed with 
 physical characteristics such as 
entropy, viscosity and electrical conductivity.
We show that certain properties of the stretched horizons
 are encoded in the quasinormal spectrum of black holes.
We compute analytically the lowest quasinormal frequency of 
a vector-type perturbation for a generic black hole with 
a translationally invariant horizon (black brane) 
in terms of the background metric components. The resulting dispersion 
relation is identical to the one obtained in the membrane paradigm
treatment of the diffusion on stretched horizons.
Combined with the Buchel-Liu universality theorem for 
the membrane's diffusion coefficient, our result means that
in the long wavelength limit the black brane
spectrum  of gravitational perturbations
exhibits a universal, purely imaginary quasinormal frequency. 
In the context of gauge-gravity duality, this provides yet
 another (third) proof of the universality of shear viscosity
 to entropy density ratio in theories with gravity duals.
\end{abstract}

\maketitle

%
%

\section{Introduction}

In the membrane paradigm approach to black holes,
the horizon is replaced by a time-like surface (the stretched horizon)
located infinitesimally
 close to the true mathematical horizon \cite{Thorne:iy}, 
\cite{Susskind:2005js}.
The stretched horizon behaves as an effective membrane endowed with 
 physical properties such as electrical conductivity and viscosity.
Furthermore, in the case of black branes 
(black holes with translationally invariant horizons) it was shown 
\cite{Kovtun:2003wp} that 
the charge density $j^0$ defined on the 
stretched horizon by the standard membrane paradigm construction 
\cite{Parikh:1997ma} obeys 
the diffusion equation  
\begin{equation}
\partial_t j^0 = D \nabla^2 j^0 
\label{deq}
\end{equation}
in the long-wavelength (hydrodynamic) limit. (The translational
 invariance of the horizon guarantees the existence of the
 hydrodynamic limit.)
The corresponding dispersion relation has the form
\begin{equation}
\omega = - i D q^2\,,
\label{disp}
\end{equation}
where $q$ is the momentum along the stretched horizon.
For a generic black brane metric, 
the  diffusion coefficient $D$ can be 
 determined explicitly in terms of the  metric components \cite{Kovtun:2003wp}.
Moreover, for the so-called shear mode of the {\it metric} fluctuation,
one can write an effective Maxwell action and an effective charge density on the stretched horizon
satisfying Eq.~(\ref{deq}) with the shear mode damping constant determined by the ambient metric.
A number of examples considered in  \cite{Kovtun:2003wp} suggested that 1) the membrane's 
shear mode damping constant was equal to $1/4\pi T$ independently of the background metric and 2)
the membrane's diffusion and the shear mode damping constants coincided, respectively, 
 with the appropriate diffusion and damping constants 
of the currents and stress-energy tensors in the dual theory computed via the AdS/CFT correspondence.
Using Einstein's equations, Buchel and Liu \cite{Buchel:2003tz} proved that the 
suggestion 1) is universally true for the class 
of metrics considered in \cite{Kovtun:2003wp}.
(An alternative proof of the universality of the membrane's coefficient employing the Lorentz boost of the black brane metric 
can be found in Section 6 of the review \cite{Son:2007vk}.)
This proof, however, cannot be viewed as the proof of the universality 
of the shear viscosity to entropy density ratio in the  
dual field theory without the proof of the suggestion 2). One of the goals of the present Essay is to supply this 
missing link by proving the  suggestion 2).
%
%
To prove it, we need either to understand why the membrane paradigm 
appears to ``know'' about the gauge/gravity duality or to derive the generic formulas of  \cite{Kovtun:2003wp} 
for the diffusion and damping constants by holographic means only, without appealing to membrane paradigm constructions.
In this Essay, we focus on the latter alternative.

More generally, our goal is to show that at least 
some of the properties of the stretched horizon are encoded
 in the quasinormal spectrum
of the corresponding black hole (brane). (For a review
 on quasinormal modes see e.g. \cite{Kokkotas:1999bd}.)
We compute analytically the 
lowest quasinormal frequency of 
a vector-type
fluctuation
 in the background of a black brane and show that 
it is of the form (\ref{disp}) with the coefficient $D$ identical
 to the diffusion coefficient computed  in \cite{Kovtun:2003wp} using
 the ``membrane paradigm'' approach. Since in the gauge/gravity duality 
the quasinormal spectrum of bulk field fluctuations is identified with the poles of the retarded 
correlators of operators dual to the fluctuations \cite{Son:2002sd}, our result
 proves the suggestion 2) and, together with 
Ref.~\cite{Buchel:2003tz} (or  Ref.~\cite{Son:2007vk}), provides yet another 
proof of the universality of the shear viscosity to entropy density ratio in 
thermal field theories in the regime described by dual classical gravity. 
(The other two proofs are Refs.~\cite{Kovtun:2004de}, \cite{Buchel:2004qq}. Clearly, all three approaches are interrelated.)


\section{Quasinormal spectrum of a $U(1)$ fluctuation in a black brane background}
\label{maxwell}

A black p-brane is represented by the metric
\begin{equation}
ds^2 = G_{tt}(r) dt^2 + G_{rr}(r) dr^2 + G_{xx}(r) \sum_{i=1}^{p}(d x^i)^2\,.
\label{metric}
\end{equation}
Such metrics typically result from a dimensional reduction of 
higher-dimensional supergravity solutions.
As a guide, one can have in mind  the near-extremal black three-brane 
solution of type II supergravity dimensionally reduced on a five-sphere 
\begin{equation}\label{exa}
  ds^2 = \frac{r^2}{R^2}( -f d t^2  + d x^2 + d y^2 + d z^2)
         + \frac{R^2}{r^2f}d r^2, \qquad
  f = 1 - \frac{r_0^4}{r^4}\,,
\end{equation}
but our discussion will be quite general.
We assume that the metric (\ref{metric}) has a translationally-invariant 
 event horizon at $r=r_0$ that extends in $p$ spatial dimensions 
parametrized by the coordinates $x^i$.
It will be convenient to introduce a dimensionless coordinate 
$u=r_0^2/r^2$ that maps the semi-infinite interval 
$r\in [r_0,\infty)$ into a finite one, $u\in [0,1]$.
The metric becomes
\begin{equation}
ds^2 = g_{tt}(u) dt^2 + g_{uu}(u) du^2 + g_{xx}(u) \sum_{i=1}^{p}(d x^i)^2\,,
\label{metricu}
\end{equation}
where the components are related to the ones in Eq.~(\ref{metric})
by trivial redefinitions \cite{Kovtun:2003wp}.
We assume that  near the horizon, i.e. in the limit $u\rightarrow 1$, 
the components $g_{tt}$, $g_{uu}$,  $g_{xx}$ behave as
\begin{subequations}
\begin{eqnarray}
  g_{tt} & = &  - (1-u)\,\Go + O(1-u)\,, \\
  g_{uu} & = & \frac{\Gu}{1-u} + O(1)\,, \\
  g_{xx} & = & O(1)\,,
\label{assumptions}
\end{eqnarray}
\end{subequations}
where $\Go$ and $\Gu$ are positive constants.
We also introduce a thermal factor function
\begin{equation}\label{factor}
f(u) = - g_{tt}(u)/g_{xx}(u)\,.
\end{equation}
The function $f(u)$ has a 
simple zero at $u=1$. 
The Hawking temperature associated with the 
background
(\ref{metricu}) is
\begin{equation}
\label{eq:membrane T}
    T=\frac{1}{4\pi} \sqrt{\frac{\Go}{\Gu}} \,.
\end{equation}
In our example of the black three-brane solution  (\ref{exa}),
the metric in the new coordinates is given by
\begin{equation}
ds^2 = 
  \frac{(\pi T R)^2}u
\left( -f(u) dt^2 + dx^2 + dy^2 +dz^2 \right) 
 +{R^2\over 4 u^2 f(u)} du^2\,, \qquad
  f = 1 - u^2\,.
\label{near_horizon_u}
\end{equation}

Consider now fluctuations of a $U(1)$ field $A_\mu (u,t,x)$ in the 
background (\ref{metricu}). This field can be viewed e.g. as a graviphoton
 of the dimensional reduction. 
Translational invariance of the horizon implies that the fluctuation 
can be taken to be proportional
 to $e^{-i \omega t + i \bf{q}\bf{x}}$, and we choose the spatial momentum
to be directed along $x\equiv x^p$.

In the gauge $A_u=0$, Maxwell's equations  
$\d_\mu \left( \sqrt{-g} \, F^{\mu\nu} \right) = 0$ for the components
$A_t(u,t,x)$ and $A_x(u,t,x)$ read
\begin{subequations}
\begin{eqnarray}
 & & g^{tt}\, \omega \, A_t' - q \, g^{xx} \, A_x' =0\,,\label{max1}\\
 & & \d_u \left( \sqrt{-g} g^{tt} g^{uu} A_t'\right) - 
 \sqrt{-g} g^{tt} g^{xx} \left( \omega q A_x + q^2 A_t \right) =0\,,
  \label{max2}\\
  & & \d_u \left( \sqrt{-g} g^{xx} g^{uu} A_t'\right) - 
 \sqrt{-g} g^{tt} g^{xx} \left( \omega q A_t + \omega^2 A_x \right) =0\,,
  \label{max3}
\end{eqnarray}
\end{subequations}
where prime denotes the derivative with respect to $u$.
 All other components of  $A_\mu (u,t,x)$ decouple, and thus can be 
consistently set to zero.



For a gauge-invariant combination $E_x = \omega A_x + q A_t$ 
 (the component of the electric field
 parallel to the brane)
 the system (\ref{max1})-(\ref{max3})
yields the following equation
\begin{equation}
  E_x'' + \Biggl[ {\wn^2 f'\over f (\wn^2 - \qn^2 f)} +
 \d_u  \log{\left( \sqrt{-g} g^{tt} g^{uu}\right) } \Biggr] E_x' +
 { (2\pi T)^2 g^{xx}\over f g^{uu}}\, \Biggl( \wn^2  - \qn^2 f  \Biggr)\, E_x =0\,,
\label{gineq}
\end{equation}
where $\wn = \omega/2\pi T$, $\qn = q/2\pi T$. 
 The differential 
equation (\ref{gineq}) has a singular point at $u=1$ with the exponents
$\alpha_{\pm} = \pm i \wn/2$ corresponding to the waves 
emerging from  and disappearing into the horizon.
 Imposing the incoming wave boundary condition 
at the horizon, one can write the solution as
\begin{equation}
E_x (u) = f^{-i \wn/2} F(u)\,,
\end{equation}
where $F(u)$ is regular at $u=1$. 
At spatial infinity, $u=0$, we impose the Dirichlet boundary condition
$E_x(0)=0$.

We are interested in computing the quasinormal spectrum 
 of the fluctuation $E_x$ subject to the 
boundary conditions stated above. Generically, we expect the spectrum 
to consist of an infinite tower $\omega_n = \omega_n (q)$ 
of the discrete complex frequencies. The lowest frequency, $\omega_0(q)$, 
can have a finite gap as $q\rightarrow 0$, or be gapless, 
$\lim\limits_{q\rightarrow 0}\omega_0 (q)=0$.
We will now show that the frequency of the vector-like fluctuation 
$E_x(u,t,x)$ is in fact gapless, and compute its value
in the limit of small $\omega$, $q$.

An analytic solution to Eq.~(\ref{gineq}) in the limit 
$\wn \ll 1$, $\qn \ll 1$ can be easily found.
Introducing a book-keeping parameter $\lambda$ and rescaling 
$\wn \rightarrow \lambda \wn$, $\qn \rightarrow \lambda \qn$,
one can obtain a perturbative solution 
in the form
\begin{equation}
E_x (u, \wn, \qn )
 = f^{-i \wn/2} \Biggl( F_0(u) + \lambda F_1(u) + O(\lambda^2)\Biggr)\,,
\end{equation}
where each of the functions $F_i(u)$ obeys an equation derived from 
 Eq.~(\ref{gineq}).
The equation for $F_0(u)$ has a generic solution
\begin{equation}
F_0 = C_0 + C_1 \int {(\wn^2 -\qn^2 f) d u\over f\sqrt{-g}
 g^{tt} g^{uu}}\,,
\label{fo}
\end{equation}
where $C_0$, $C_1$ are integration constants.
The integral in Eq.~(\ref{fo}) is logarithmically divergent at $u=1$. 
Since  
by construction  $F_0$ is a regular function, we must put $C_1=0$.
The function $F_1(u)$ obeys an inhomogeneous equation whose 
 regular at $u=1$
solution is given by
\begin{equation}
F_1 (u) = - {i \wn C_0\over 2} \log{f} - {i C_0 \sqrt{-g(1)}f'(1)\over
2 \wn \gamma_0 \gamma_u}  \int {(\wn^2 -\qn^2 f) d u\over f\sqrt{-g}
 g^{tt} g^{uu}}\,.
\label{f1}
\end{equation}
Given these explicit solutions, we use the Dirichlet condition
$E_x(0)=0$  to obtain the following equation for $\wn$
\begin{equation}
\wn - i  \qn^2  { \sqrt{-g(1)}f'(1)\over
2  \gamma_0 \gamma_u}  \int_{0}^{1} { d u\over \sqrt{-g}
 g^{tt} g^{uu}}\, + O (\wn^2) =0\,.
\label{d1}
\end{equation}
Solving  Eq.~(\ref{d1}) 
for $\omega$ and taking into account Eqs.~(\ref{assumptions}), (\ref{factor}), 
we find the lowest quasinormal frequency
\begin{equation}
\omega = - i D q^2 + O(q^4)\,,
\label{q1}
\end{equation}
where the diffusion constant is given by 
\begin{equation}
D = {\sqrt{-g(1)} \over g_{xx}(1) \sqrt{- g_{tt}(1)g_{uu}(1)}}
 \int_{0}^{1} d u { - g_{tt}(u) g_{uu}(u)\over \sqrt{-g(u)}}\,.
\label{dc}
\end{equation}
%
The formula (\ref{dc}) coincides precisely with the result obtained 
using the 
membrane paradigm approach to the diffusion on stretched horizons 
\cite{Kovtun:2003wp}.

As a simple application, consider computing a $U(1)$ charge diffusion constant
 in a $p+1$-dimensional conformal field theory at strong coupling. The dual gravitational
AdS-Schwarzschild background is given by the metric 
\begin{equation}
ds^2 = \frac{r^2}{R^2}( -f d t^2  + d \vec{x}^2_p )
         + \frac{R^2}{r^2 f}d r^2, \qquad
  f = 1 - \frac{r_0^{p+1}}{r^{p+1}}\,.
\end{equation}
Changing variables to $u=r_o^2/r^2$ and using Eq.~(\ref{dc}) we
 find\footnote{Formula (\ref{diffco}) was independently obtained in Ref.~\cite{Kovtun:2008kx}.}
\begin{equation}
D = \frac{p+1}{p-1}\, \frac{1}{4\pi T}\,.
\label{diffco}
\end{equation}
For strongly coupled CFT$\,_{2+1}$,  CFT$\,_{3+1}$, CFT$\,_{5+1}$,  Eq.~(\ref{diffco}) gives $D=3/4\pi T$, 
$D=1/2\pi T$ and  $D=3/8\pi T$, respectively
which coincides with the R-charge diffusion constants found by 
computing the poles of the retarded correlators in AdS/CFT \cite{Policastro:2002se}, \cite{Herzog:2002fn}.
Other applications of the result (\ref{dc}) in AdS/CFT context can be found in \cite{Kovtun:2003wp}, \cite{Myers:2007we}.

\section{The shear mode gravitational perturbation and universality}

Gravitational perturbations $h_{\mu\nu}$ of the black brane background metric (\ref{metricu}) 
can be classified according to their transformation properties into 
the scalar (sound mode), vector (shear mode)
and tensor types \cite{Kovtun:2005ev}. The vector type fluctuations $h_{ty}(u,t,x)$, $h_{xy}(u,t,x)$, $y\neq x$
can be treated as components of an effective $U(1)$ field $A_t = (g_{xx}^{-1})\, h_{ty}$, $A_x = (g_{xx}^{-1})\, h_{xy}$
satisfying the equation 
$$
\partial_\mu \left( \sqrt{-g} \, g_{xx}^{\frac{p}{p-1}} \, F^{\mu\nu} \right) = 0
$$ 
in the black brane background \cite{Kovtun:2003wp}. 
Thus the computation of the shear mode lowest quasinormal frequency and the damping constant 
is essentially identical to the one presented in Section \ref{maxwell}. For $\omega\ll T$, $q\ll T$, the 
frequency is given by $\omega = - i {\cal D} \, q^2 + O(q^4)$, where the shear mode damping constant is given by
 \begin{equation}
{\cal D} = {\sqrt{-g(1)} \over \sqrt{- g_{tt}(1)g_{uu}(1)}}
 \int_{0}^{1} d u { - g_{tt}(u) g_{uu}(u)\over g_{xx}(u) \sqrt{-g(u)}}\,.
\label{dampc}
\end{equation}
Again, the quasinormal mode 
coefficient ${\cal D}$ coincides with the membrane's shear mode coefficient 
obtained from the membrane paradigm \cite{Kovtun:2003wp}.
Buchel-Liu theorem \cite{Buchel:2003tz} or the argument presented in \cite{Son:2007vk}) show that 
the  membrane's shear mode coefficient \ref{dampc} is universal and equal to 
 to $1/4\pi T$ for all metrics of the form  (\ref{metricu}). 
We immediately conclude that all black brane metrics 
 of the form  (\ref{metricu}) possess a universal, 
purely imaginary gravitational quasinormal frequency 
\begin{equation}
\omega = -i q^2/4\pi T + O(q^4)\,,
\label{universal}
\end{equation}
where $T$ is the Hawking temperature
associated with the metric (\ref{metricu}).
Moreover, since the gauge/gravity duality dictionary  \cite{Son:2002sd}, 
\cite{Kovtun:2005ev}  identifies the quasinormal frequency 
(\ref{universal}) as the pole $\omega = - i \eta q^2 / s T + ...$ 
of the correlator of the shear components of the stress-energy tensor 
in a finite temperature field theory dual to the background (\ref{metricu}), the universality of the quasinormal mode 
implies the universality of the  ratio of the shear viscosity $\eta$ to the entropy density $s$ in the dual theory.

\section{Discussion}

We have computed analytically the lowest quasinormal frequencies of the electromagnetic and
the shear mode gravitational fluctuations in a  black brane background. 
The spectrum of gravitational perturbations exhibits a universal, purely imaginary quasinormal frequency.
In the context of  the gauge/gravity duality, this implies the universality of the viscosity to entropy 
density ratio in the dual finite-temperature field theory. Transport coefficients of field theories  
with gravity duals are completely determined (in the regime where the gravity dual description is valid) 
by the lowest quasinormal frequencies of the dual gravity backgrounds.
On the one hand, at this point we can completely ignore any connection with the black hole membrane paradigm and
concentrate of computing the zero-frequency limits of relevant quasinormal spectra.
The simplest of such calculations is presented in this Essay. Along the same lines, one can show that the 
tensor gravitational mode in a black brane background is not gapless (as expected from the absence
 of a relevant conserved current in a dual theory). Computing the sound mode spectrum for a generic 
gravitational background would be of great interest as it would provide formulas for the bulk viscosity 
and the speed of sound similar to the expression for the charge diffusion coefficient in Eq.~(\ref{dc}).
Finding such formulas would be more difficult since the diagonal metric perturbations couple to fluctuations of 
other background fields.

On the other hand, the fact that the membrane's diffusion coefficients are encoded in the quasinormal
spectrum is intriguing. In particular, universality of the lowest gravitational mode may reflect 
horizon properties independently of any AdS/CFT interpretation. As noted in \cite{Kovtun:2003wp},
the ``shear viscosity'' of a stretched horizon of an asymptotically flat Schwarzschild black hole computed 
in the ``old'' membrane paradigm framework \cite{Damour}, 
\cite{Parikh:1997ma}, still obeys $\eta/s = 1/4\pi$, despite the apparent absence 
of a holographic dual theory and the fact that the ``bulk viscosity'' and the specific heat are negative.
Perhaps new approaches \cite{Hod:2006jw,Saremi:2007dn, Fujita:2007fg, Konraad, Bhattacharyya:2008jc, 
Bhattacharyya:2008xc,Bhattacharyya:2008ji,Eling:2008af} will help to clarify these issues.

From a ``holographic'' perspective, it is clear that the stretched horizon is a slice 
of an asymptotically AdS background located in the ``deep infrared''.  
The imaginary part of the retarded correlators (whose zero-frequency limit gives transport coefficients via Green-Kubo
 formulas) is independent of the background radial coordinate \cite{Son:2002sd}. 
It is thus conceivable that the membrane paradigm and holography are intimately related to each other.
However, further work is required to quantify these philosophical sentiments.

\vspace{0.25cm}

\noindent
{\bf Acknowledgments}

\noindent
Research at Perimeter Institute is supported by the Government of Canada 
through Industry Canada and by the Province of Ontario through 
the Ministry of Research \& Innovation.

\vspace{0.5cm}

\noindent
{\bf Note added:}
{\small The original version of this note was submitted as an Essay to 
2006 Gravity Research Foundation competition (see also Theory Canada II 
conference http://pirsa.org/06060018). It has never appeared in print.
In 2007 and 2008, respectively, I was informed by Anshuman Maharana and by 
Hong Liu that they independently obtained the same analytic result for the 
lowest quasinormal frequency of the vector fluctuation.}

\end{document}